\documentstyle[11pt]{article}

\title{\bf Leptoproduction of nucleons
in the cumulative region}

\author{M.A.Braun, V.V.Vechernin\\
Dep. High-Energy physics,
S.Petersburg University, 198904 S.Petersburg, Russia,\\
 and B.Vlahovic\\
NCCU, Durham, USA}
\date{}
\def\beq{\begin{equation}}
\def\eeq{\end{equation}}

\begin{document}
\maketitle
\medskip
\vspace{1 cm}

{\bf Abstract}

Leptoproduction of nucleons into the backward hemisphere on
nuclear targets is studied  at relativistic subasymptotic
energies and momenta. Spins are neglected. The relativistic
internucleon potential is extracted from the appropriate
photoproduction data. Different
production mechanisms are shown to work together and interfere.
Calculations show that whenever rescattering is possible it gives
the bulk of the contribution except at very high $Q^2$.
The Weizsaecker-Williams approximation is found to generally reproduce
only a small part of the total cross-section. Comparison with the
data for A(e,e'p) reaction at E=2.4 GeV shows a reasonable agreement.

\section{Introduction}
Production of particles off nuclei in the cumulative domain $x>1$
presents a great interest, since it is related to the
nuclear structure at small distances, where one may expect
formation of multiquark states and even droplets of a cold quark-
gluon plasma. Staying within the most conservative picture 
of the nucleus consisting of nucleons interacting with some 
potentials,  cumulative particle production allows to study
the high-momentum asymptotics of these potentials and, which is
of especial importance,  multi-nucleon forces, otherwise
hidden due to the low probability for several nucleons to be
located at the same point. Experimental
data demonstrate that the rate of cumulative production is very weakly
dependent on the initial energy, starting from  several GeV's.
This allows to study  cumulative production at comparatively 
low energy facilities, provided  their luminosity is high
enough to observe cross-sections dramatically falling as $x$ grows
beyond the threshold $x=1$.

There has never been shortage of theoretical models to 
describe cumulative 
production off nuclei. All proposed models can be roughly 
divided into three categories. The first (also in time) is to
ascribe the production to the presence 
of states with a nuclear density much higher than on the average
in the initial nucleus [1,2]. 
Interpretation of these states may be different: just several
nucleons at the same point ("many-body correlations"[3]) or
multi-quark states [4]). The second mechanism  assumes that
such high-density states are rather formed in the course of
the collision ("compressed tube" models [5]). Finally one can
assume that the high nuclear densities are irrelevant and
the cumulative particles are produced as a result of simple
rescattering [6].

Microscopic studies, both in the nucleon and quark pictures,
have shown that in reality all three mechanisms work together
in the cumulative production. Moreover their contributions
interfere, so that the total rate is not just the sum of them [ 7 ].
It is of most importance that the contribution from the rescattering
depends differently on the target properties that the other two.
This prevents simple estimates of the cross-sections on heavy nuclei
based on those with only few nucleons, as suggested in [ 8 ].
Only at sufficiently high momentum transferred to the nucleus, 
of all contributions, only the first one remains, corresponding 
to the few-nucleon correlations or multi-quark states inside
the nucleus. From the theoretical point of view, it is the
most interesting part, since it reveals the inner structure of the
nucleus at short distances. However to see it clearly against
the background of other components, one has to select events with 
high enough $Q^2$, which drastically diminishes already
small  cross-sections. Also the exact values of $Q^2$ at which
one can neglect the contributions from other two mechanisms
are not known {\it apriori}.

For this reason it is important to carry out calculations of
cumulative production at finite $Q^2$ which take into account
all three above mentioned mechanisms. In this
way one expects to obtain predictions for the inclusive
cross-sections, integrated over all $Q^2$, and thus directly
related to the bulk of experimental data. Moreover, studying
the relative weight of the contributions at different $Q^2$
one may determine the minimum $Q^2$ value, starting from which
rescattering and compressed tube contributions can be safely
neglected.

In this paper we do these calculations in the framework of 
the nucleon degrees of freedom. Estimates have shown that this
description gives reasonable results at $x$ not too close to the
cumulative threshold $x=n$ for $n$ nucleons which interact at
close distance from each other, provided one uses the correct 
relativistic kinematics [ 3 ]. We limit ourselves with the cumulative region
$1<x<2$, easiest  from both experimental and theoretical points of view.
In this region it is commonly assumed that one has to study  hard 
interactions of only two nucleons inside the nucleus.
 Since inclusion of relativistic spins presents
serious difficulties for nuclei with $A>2,3$, we simplify our approach 
by taking  nucleons scalar. This approximations may certainly
change results by a factor of order 2$\div$3. However in the cumulative
region as $x$ grows from 1 to 2 the cross-sections fall by several orders
of magnitude. On this scale a factor of order unity plays a secondary role.
In any case there is no reason to expect that the results obtained with 
scalar nucleons will have a different $x$, $p_\perp^2$ and $A$ dependence
as compared to the full spinor case, and it is this dependence which
is of primary theoretical interest. 

The paper is organized as follows. In Section 2 we present our basic 
formulas for the structure function and solve the most difficult task of 
extracting
the rescattering contribution. In Section 3 we discuss how to pass from
the structure function to the production rates.
Our final results
are presented  in Section 4. Section 5 contains some discussion and
conclusions.
Appendix is devoted to
certain non-trivial details of numerical calculations.

\section{Structure functions in the interval $1<x<2$}
\subsection{Cross-sections and kinematics}
We shall be interested in the inclusive leptoproduction of
nucleons on nuclear targets in the cumulative kinematical domain.
However it will be convenient to start with a simpler reaction
\beq 
e+A\to e+X.
\eeq
Its cross-section can be expressed via nuclear structure
functions in the standard manner
\beq
\frac{d\sigma}{dE'd\Omega'}=\sigma_{Mott}\frac{1}{4\pi M}
\left(\frac{M^2}{qP}F_2(x,Q^2)+2\tan^2\frac{\theta}{2}F_1(x,Q^2)\right),
\eeq
where
\beq
\sigma_{Mott}=4\frac{\alpha_{em}^2}{Q^4}(E')^2\cos^2 \frac{\theta}{2}.
\eeq
In our normalization the structure functions are related to the imaginary
part of the forward amplitude for the elastic $\gamma^*-A$ scattering as
\beq
 W_{\mu\nu}=\left(-g_{\mu\nu}+\frac{q_{\mu}q_{\nu}}{q^2}\right)F_1(x,Q^2)+
\frac{1}{qP}\left(P_\mu-q_\mu\frac{qP}{q^2}\right)
\left(P_\nu-q_\nu\frac{qP}{q^2}\right) F_2(x,Q^2),
\eeq
where $q$ and $P=Ap$ are the momenta of the photon and nucleus and, as 
usual,
\beq
Q^2=-q^2,\ \ x=\frac{Q^2}{2qp}.
\eeq
Note that we define $x$ respective to  $\gamma^*+N$ scattering, so that
\beq
0\leq x\leq A
\eeq
The region $x>1$ is cumulative.

The standard way to find the structure functions from $W_{\mu\nu}$ is to 
choose
a coordinate system in which $q_+=q_y=p_{\perp}=0$ ("theoretical"). 
Marking the components of 
vectors and tensors in this system  with bars,  one finds
\beq
F_1(x,Q^2)=\bar{W}_{yy},\  \  F_2(x,Q^2)=\frac{Q^2}{2Axp_+^2}\bar{W}_{++}.
\eeq
We take the nucleon at rest: ${\bf p}=0$. To study leptoproduction
of cumulative nucleons we shall have to use another system ("lab")
in which the incoming lepton with momentum $l$ 
is moving along the $z$-axis in the opposite direction, so that
neglecting its mass $l_+=0$.
These two systems are related by a spatial rotation, which can be split
into two ones.
Starting from the lab. system, the first rotation is to put the
$y$-component of
${\bf q}$ (or ${\bf l'}$) to zero. This is just a rotation in the
$xy$ plane by the azimuthal angle $\phi$:
\beq
q'_x=q_x\cos\phi+q_y\sin\phi,\ \ q'_y=-q_x\sin\phi+q_y\cos\phi=0,\ \
q'_z=q_z.
\eeq
The second rotation in $xz$ plane is to put $q_+=0$, that is $q_z=-q_0$:
\beq
q_z''=q'_z\cos\phi_1+q'_x\sin\phi_1=-q_0,\ \
q''_x=-q'_z\sin\phi_1+q'_x\cos\phi_1.
\eeq
The first relation determines the value of angle $\phi_1$ through the given
 value of the transferred energy $q_0=E-E'$:
\beq
\cos\phi_1=\frac{1}{{\bf q}^2}(-q_0q_z\pm Qq'_x).
\eeq
The two signs correspond to the two possibilities of directing $q_x$
parallel or antiparellel to the $x$-axis. 

To find the inclusive cross-section for the leptoproduction of
cumulative nucleons we shall integrate the cross-section (2)
over $E'$ and $\Omega'$ , present both structure functions in
(2) as integrals over intermediate particles and then lift the integration
over the momenta of the observed nucleon.
The only point
to have in mind is that the structure function and the inclusive
cross-section
are to be calculated in different reference systems, theoretical and lab
respectively. Since both $Q^2$ and $qp$ will vary in the resulting
expression for the production rate, it will be more convenient to
characterize the cumulative region directly in terms of the momentum 
$p_1$ of the observed nucleon in the lab system. We shall discuss the
corresponding kinematics in Section 3.

The method we use to find the
cross-section for the inclusive leptoproduction of nucleons
from the inclusive leptoproduction of leptons is discussed and
justified in Appendix 1. There one can also find the expression for
the double inclusive cross-section for the process in which both
the lepton and cumulative nucleon are observed. This cross-section
can be found from (2) only averaged over the azimuthal
directions of the observed nucleon.
\subsection{Hard interactions between two nucleons}
In the interval $n-1<x<n$ the incoming virtual photon has to interact with
at least $n$ nucleons. Since the cross-section results rapidly diminishing
with the number of interacting nucleons, our first approximation will be to
take into account  only the interaction with $n$ nucleons in this interval.
Then at $1<x<2$ the incoming photon is to interact with two and only two
nucleons (approximation of "pair correlations" for the 
short-range
part). The two interacting nucleons
will generally have large momenta ${\bf k}$ and ${\bf q-k}$, wheras the
rest intermediate particles will be slow. Their momenta will be of the order
of the standard nuclear ones, to be neglected wherever possible.
Separating the large momentum part from the low momentum one we
can present the total contribution to the ampliude for the reaction 
$\gamma^*+A\to N(p_1)+N(p_2)+X$
as a sum
of four diagrams, shown in Fig. 1, $a-d$, where the thin lines represent
slow nucleons, thick ones represent fast or highly virtual nucleons and
thick dashed lines represent the interaction between the nucleons with
high momentum transfer.

An important point is that the high-momentum part responsible for the
production of the cumulative nucleons involves both short- and long-range
distances in the nucleus, the latter corresponding to the rescattering
contribution. Also to note is that the short-range contribution comes
both from
diagrams $a$ and $b$ and from diagrams $c$ and $d$. The first pair of
diagrams corresponds to the presence of the correlated pair inside the
initial
nucleus before the collision. Ideologically this is close to the
"flucton" model,
which was used to explain the cumulative phenomena immediately after
their observation. The second pair of diagrams, $c$ and $d$, correspond
however to correlations created in the course of the collision. This is
in line
with mechanisms proposed in so-called "hot' models, like the 
"compressed tube model". As we observe, inspection of Feynman diagrams
shows that in fact all different mechanisms proposed so far to explain
the cumulative phenomena coexist simultaneously. Moreover, as we shall
discover presently, their contributions to the cross-section are not added
incoherently, but interfere, which leads to important cancellations.
Our  task will be to separate the long-range rescattering contribution,
and to take into account the mentioned
interference, which we shall do following [ 6 ] .

The imaginary part of the forward scattering  amplitude corresponding
to Fig. 1 is given by
\[
W_{\mu\nu}=\frac{1}{2}f^2(Q^2)
\int \frac{d^3p_1}{(2\pi)^32p_{10}}\frac{d^3p_2}{(2\pi)^32p_{20}}
\prod_{j=1}^2\frac{d^4k_j}{(2\pi)^4}\frac{d^4k'_j}{(2\pi)^4}\]\beq
(2\pi)^4\delta^4(k_1+k_2-k'_1-k'_2)(2\pi)^4\delta(p_1+p_2-q-k_1-k_2)
H_{\mu}(H'_{\nu})^*T(k_1,k_2|k'_1,k'_2),
\eeq
where the notations are clear from Fig. 1. In particular $f(Q^2)$ is the
effective nucleon form-factor;
blob $T$ which includes the
propagators of the nucleons 1 and 2 describes the nuclear recoil particles.

Factors $H$ and $H'$ depending on $k_j$ and $k'_j$, $j=1,2$ respectively,
correspond to the high-momentum part of the amplitude.
Their denominators in the virtual nucleon
propagators in Fig. 1 $a$ and $b$ are respectively
\[m^2-(2p-p_1)^2,\ \ {\rm and}\ \ m^2-(p_1-q)^2.\]
Obviously they cannot vanish, since a stable nucleus cannot decay into a real
nucleon and some other particles. The denominator in Fig. 1 $c$ and $d$ is
\[
m^2-(q+k_1)^2=
\Big[E({\bf k_1+q})-E({\bf k_2})+E({\bf p_1})+E({\bf q-p_1+k_1+k_2})\Big]
\]\beq
\Big[E({\bf k_1+q})+E({\bf k_2})-E({\bf p_1})-E({\bf q-p_1+k_1+k_2})\Big].
\eeq
The first factor is the difference between
the initial and final energies in the decay
\[
N({\bf k_2})\to\bar{N}({\bf -k_1-q})+N({\bf p_1})+N({\bf q-p_1+k_1+k_2})\Big]
\]
and cannot vanish due to stability of the nucleon. For this reason we can
drop small momenta $k_{1,2}$ in it.
The second factor is  the difference between the initial and final
energies in the rescattering
\[
N({\bf k_1+q})+N({\bf k_2})\to N({\bf p_1})+N({\bf q-p_1+k_1+k_2})
\]
and evidently may vanish. 
Taking into account the smallness of momenta $k_1$ and $k_2$
we can rewrite it as
\beq
E({\bf q})+m-E({\bf p_1})-E({\bf q-p_1})+\frac{{\bf k_1q}}{E({\bf q})}-
\frac{{\bf (k_1+k_2,q-p_1)}}{E({\bf q-p_1})}.
\eeq
So finally the denominator (12) takes the form
\beq
m^2-(q+k_1)^2=a(u+\alpha-i0),
\eeq
where
\beq
a=E({\bf q})+E({\bf p_1})+E({\bf q-p_1})-m,
\eeq
\beq
u=E({\bf q})+m-E({\bf p_1})-E({\bf q-p_1})
\eeq
and
\beq
\alpha=\frac{{\bf k_1q}}{E({\bf q})}-
\frac{{\bf (k_1+k_2,q-p_1)}}{E({\bf q-p_1})}.
\eeq

Thus approximating $k_1=k_2=p$ everywhere except in the denominator (14) we
find the high-momentum factor as
\beq
H_{\mu}=K_{\mu}^{(a)}\frac{v(p_1-p)}{m^2-(2p-p_1)^2}+
K_{\mu}^{(b)}\frac{v(p_1-q-p)}{m^2-(q-p_1)^2}+
K_{\mu}^{(c)}\frac{v(p_1-p)+v(p_1-q-p)}{a(u+\alpha-i0)}.
\eeq
We are only interested in $+$ and $y$ components of $H$. Factors $K_+$ and
$K_y$ are identical for $H$ and $H'$ and in our system they are
\beq
K^{(a)}_+=2(2p_+-p_{1+}),\ \ K^{(b)}_+=2p_{1+},\ \ K^{(c)}_+=2p_+
\eeq
and
\beq
K^{(a)}_y=-K^{(b)}_y=-2p_{1y},\ \ K^{(c)}_y=0.
\eeq

To move further we first integrate over the zero components of $k_j$ and
$k'_j$.
The $\delta$ functions which correspond to the conservation of
energy  have the form
\beq
\delta(k_{10}+k_{20}+P'_0-P_0)\delta(k_{10}+k_{20}+P'_0-P_0)
\delta(p_{10}+p_{20}+P'_0-q_0-P_0).
\eeq
Here $P_0$ and $P'_0$ are the energies of the target nucleus and its
debris respectively.
Using the first two $\delta$-function we integrate over the zero
components of
the two active nucleons 1 and 2. In coordinate space this integration
puts both active nucleons at the same time.
If we take into account only poles of their
propagators and free nucleons for the debri, it produces the standard 
denominators of the wave function of the target nucleus. 

Now we make the decisive approximation and neglect small energies of
the nuclear debri
as compared to fast particles in the last $\delta$-function in (21)
taking it as
\beq
\delta(p_{10}+p_{20}-q_0-2m).
\eeq
This allows to integrate blob $T$ over the energy $P'_0$ of the debri,
which puts
initial and final active nucleons at the same time point. As a result $T$
transforms
into the nuclear $\rho$-matrix for a pair of active nucleons:
\beq
\int \frac{dk_{20}}{2\pi}\frac{dk'_{20}}{2\pi}\frac{dP'_0}{2\pi}
T(k_1,k_2|k'_1,k'_2)=
A(A-1)\rho({\bf k_1,k_2|k'_1,k'_2}).
\eeq

It is convenient to pass to the $\rho$ matrix in the coordinate space presenting:
\[
(2\pi)^3\delta^3(k_1+k_2-k'_1-k'_2)\rho({\bf k_1,k_2|k'_1,k'_2})\]\beq=
\int \prod_{j=1}^2d^3r_jd^3r'_j\rho({\bf r_1,r_2|r'_1,r'_2})
e^{i({\bf k_1r_1+k_2r_2-k'_1r'_1-k'_2r'_2})}
\eeq
and write the imaginary part of the forward scattering amplitude in the
form
\[
W_{\mu\nu}=\frac{1}{2}A(A-1)f^2(Q^2)\int d\tau\int \prod_{j=1}^2
\frac{d^3k_jd^3k'_jd^3r_jd^3r'_j}{(2\pi)^6}\]\beq
\rho({\bf r_1,r_2|r'_1,r'_2})
e^{i({\bf k_1r_1+k_2r_2-k'_1r'_1-k'_2r'_2})}H_{\mu}(H'_{\nu})^*,
\eeq
where 
\beq
d\tau=\frac{d^3p_1}{(2\pi)^32p_{10}}\frac{d^3p_2}{(2\pi)^32p_{20}}
(2\pi)^4\delta(p_1+p_2-q-2p)
\eeq
is the standard phase volume for the virtual photo-production on a pair
of nucleons at
rest.

\subsection{Short and long range parts}
 We start with the component $W_{yy}$. As follows from (20) it does not
contain
any terms singular at $k_{j}=k'_{j}=0$, $j=1,2$ and so can be taken at
this point.
Then integration over  $k_{j}$ and $k'_{j}$, $j=1,2$ gives four
3-dimensional $\delta$-
functions which will put coordinates $r_{j}=r'_{j}=0$, $j=1,2$:   
\beq
W_{yy}=\frac{1}{2}A(A-1)f^2(Q^2)\rho_0\int d\tau H_{y}^2 ,\ \
\rho_0\equiv\rho(0,0|0,0).
\eeq
Obviously this corresponds to a pure short range mechanism, both direct and
spectator.

Passing to $W_{++}$ we split $H_+$ into parts singular and non-singular at
$k_{j}=0$, $j=1,2$
\beq
H_+=\tilde{H}_++\frac{B}{u+\alpha-i0},
\eeq
where $\tilde{H}_+$ is a non-singular part and according to (18)
\beq
B=2p_+\frac{v(p_1-p)+v(p_1-q-p)}{E({\bf q})+E({\bf p_1})+E({\bf
q-p_1})-m}.
\eeq
Doing the same for $(H'_{+})^*$ we find
\beq
H_+(H'_+)^*=\tilde{H}_{+}^2+2B\tilde{H}_+{\rm Re}\,\frac{1}{u-i0}+B^2
\frac{1}{u+\alpha-i0}\frac{1}{u+\alpha' +i0},
\eeq
where we have put $k_{j}=k'_{j}=0$, $j=1,2$, wherever possible.

The first two terms in (30) do not depend on $k_{j},k'_j$, $j=1,2$
and so integration over these leads to a short-range contribution similar
to (27)
\beq
\tilde{W}_{++}=\frac{1}{2}A(A-1)f^2(Q^2)\rho_0
\int d\tau \left(\tilde{H}_{+}^2+2BH_+{\rm Re}\,\frac{1}{u-i0}\right).
\eeq

To treat the singular contribution we first symmetrize it in primed and
non-primed
momenta
\[X\equiv\frac{1}{u+\alpha-i0}\frac{1}{u+\alpha' +i0}\to
\frac{1}{2}\left(\frac{1}{u+\alpha-i0}\frac{1}{u+\alpha' +i0}+
\frac{1}{u+\alpha+i0}\frac{1}{u+\alpha' -i0}\right).\]
Next we use
\[\frac{1}{u+\alpha-i0}=\frac{1}{u+\alpha+i0}+2\pi\delta(u+\alpha)\]
to find
\[
X=\frac{1}{2}\Big(\frac{1}{u+\alpha+i0}\frac{1}{u+\alpha' +i0}+
2\pi i\delta(u+\alpha)\frac{1}{u+\alpha'+i0}\]\[
+\frac{1}{u+\alpha-i0}\frac{1}{u+\alpha' -i0}
-2\pi i\delta(u+\alpha)\frac{1}{u+\alpha'-i0}\Big)\]\beq=
{\rm Re}\, \frac{1}{(u+\alpha-i0)(u+\alpha'-i0)}+
2\pi^2\delta(u+\alpha)\delta(u+\alpha').
\eeq

The first term is non-singular at $k_j=k'_j=0$, $j=1,2$ and can be
rewritten as
${\rm Re}\, 1/(u-i0)^2$. The second  reduces to
$2\pi^2\delta(u)\delta(\alpha-\alpha')$. Taking into account that
$k_1+k_2=k'_1+k'_2$ we
find that
$\alpha-\alpha'=(\bf{q,k_1-k'_1})/E({\bf q}) $, so that finally
\beq
X={\rm}\,\frac{1}{(u-i0)^2}+2\pi^2 E({\bf
q})\delta(u)\delta(\bf{q,k_1-k'_1}).
\eeq

The first term in (33) does not depend on $k_j=k'_j$, $j=1,2$ and its
integration is similar to
the previous terms. It also corresponds to a short-range contribution and
summing with (31)  forms the complete short-range  part of $W_{++}$: 
\beq
W^{(s)}_{++}=\frac{1}{2}A(A-1)f^2(Q^2)\rho_0
\int d\tau \left(\tilde{H}_{+}^2+2B\tilde{H}_+{\rm Re}\,\frac{1}{u-i0}+
B^2{\rm Re}\,\frac{1}{(u-i0)^2}\right).
\eeq

The second term in (33) describes the long-range mechanism corresponding
to real rescattering.
It contains non-trivial dependence on small nuclear momenta. To do the
integrations over
$k_j=k'_j$, $j=1,2$ we direct ${\bf q}$ along the $z$-axis. Then
integrations over
$k_{1\perp}, k'_{1\perp},k_2, k'_2$ and the corresponding coordinates
are done as before and
the only non-trivial integrations left are over $k_{1z}$ and $k'_{1z}$.
We find the long-range
part of $W_{++}$ as
\[
W_{++}^{(l)}=\frac{1}{2}A(A-1)
\int d\tau 2\pi^2\frac{E({\bf q})}{|{\bf q}|}B^2\delta(u)
\int \frac{dk_{1z}}{2\pi}dz_1dz'_1
e^{ik_{1z}(z_1-z'_1)}\rho(0_{\perp},z_1;0|0_{\perp},z'_1;0)\]\beq
=2\pi^2l\frac{E({\bf q})}{|{\bf q}|}\int d\tau B^2\delta(u),
\eeq
where
\beq
l=\frac{1}{2}A(A-1)\int dz\rho(0_{\perp},z;0|0_{\perp},z;0)
\eeq
has a meaning of the average dimension of the nucleus, travelled by the
nucleon
in rescattering. Note that due to energy conservation  $u$ and $a$  can
be rewritten as
\beq
u=E({\bf q})-q_0-m,\ \ a=E({\bf q})+q_0+m
\eeq 
and do not depend on integration variables in (35). So one can take
$\delta(u)$ out of the
integral (35). This leads to our final formula for the long-range
contribution
\beq
W_{++}^{(l)}=2\pi^2l f^2(Q^2)\delta(E({\bf q})-q_0-m)
\frac{E({\bf q})}{|{\bf q}|}\int d\tau B^2.
\eeq
Note that it has a $\delta$-like behaviour in $x$. Indeed
\[\delta(E({\bf q})-q_0-m)=2(q_0+m)\delta(E^2({\bf q})-(q_0+m)^2)\]\[=
2(q_0+m)\delta(Q^2-2mq_0)=\delta(x-1)\frac{Q^2+2m^2}{2mQ^2}.\]
So in fact the rescattering does not give any contribution to the structure
function in the  cumulative region. However it does contribute to 
cumulative particle production.

Note finally that our nuclear $\rho$-
matrix is defined in the relativistic normalization. Its relation to 
the non-relativistic one is given by
\[\rho=\frac{A}{2m}\rho_{NR}.\]


\section{Cumulative nucleon production}
We shall be interested in the observation of a nucleon of
momentum $p_1$ produced in the reaction
\beq
l+A\to N(p_1)+X,
\eeq
where $l$ is the incoming lepton of momentum $l$ and $A$ is the target
nucleus at
rest. We shall direct $\bf l$ antiparallel to the $z$-axis to have the
system close
to the theoretical system in which $q_+=0$. With such a choice
non-cumulative
nucleons will
be produced in the {\it backward} hemisphere, whereas cumulative
ones will
be observed also in the forward hemisphere. The exact non-cumulative
kinematical
region in the lab. system can be easily established in light-cone 
variables, using the
fact that in this system  $l_+=0$. Let $p_{1+}=x_1p_+$. Obviously
in the non-cumulative region $0<x_1<1$.
Conservation of the "-" component of the momentum (light-cone energy)
requires
\beq
l_-+p_-=p_{1-}+p'_-,
\eeq
where $p'$ is the momentum of the undetected particles. The 
minimal mass of the latter is zero. So in terms of initial energy $E$
we get a relation
\beq
\frac{1}{\sqrt{2}}(2E+m)\geq \frac{m^2+p_{1\perp}^2}{2x_1p_+}+
\frac{p_{1\perp}^2}{2(1-x_1)p_+},
\eeq
which leads to the condition
\beq
p_{1\perp}^2\leq x_1(1-x_1)m(2E+m)-(1-x_1)m^2
\eeq
Together with the relation between $x_1,p_{\perp}$ and $p_{1z}$:
\beq
p_{1z}=\frac{1}{2}m\left(x_1-\frac{m^2+p_{1\perp}^2}{x_1m^2}\right)
\eeq
Eq. (42) determines the non-cumulative region in the $p_{1z},p_{1\perp}$
plane. The cumulative region for the pair correlation is determined
by the corresponding energy conservation relation
\beq
l_-+2p_-=p_-+p'_-,
\eeq
where now the minimal mass of  undetected particles equals $m$.
In terms of $x_1$ and $p_{1\perp}$ this gives a relation
\beq
p_{1\perp}^2=x_1(2-x_1)m(E+m)-m^2,
\eeq
which together with (43) determines the cumulative region for the
pair correlation. For the incident energy $E=6$ GeV in the backward
hemisphere
this region is limited by the solid line in Fig. 2
The dotted line 
limits the region corresponding to the  rescattering discussed in the
previous section. 
  
As mentioned shall find the rate of cumulative nucleon production by
using our
results for the structure
functions. Putting them into the integrated  cross-section  (2)
we then
lift the integration over the momentum $p_{1}$ of the intermediate real
nucleon.
However we have to take into account that our expressions for the
structure functions
have been obtained in a special coordinate system $q_+=q_y=0$ different
from
the lab. system for the reaction (1). This implies that we have to
reexpress all
momenta in the system $q_+=q_y=0$ via the momenta in the lab. system.

Turning to our expressions for the structure functions, we see that in
most places
the vectors appear in the form of scalar products or invariant integration
measures,
which do not change under rotations. The only quantities which
change their form under rotations are
vectors $K$, in which we shall have to
express the $+$- and $y$-components of  vector $p_1$ in the system
$q_+=q_y=0$ by its components in the lab.system:
\beq
p_{1y}\to \bar{p}_{1y}=-p_{1x}\sin\phi+p_{1y}\cos\phi,\ \ 
p_{1z}\to\bar{p}_{1z}= p_{1z}\cos\phi_1+(p_{1x}\cos\phi+p_{1y}\sin\phi)
\sin\phi_1
\eeq 
and keeping the same $p_0$.

Thus we get for the inclusive cross-section to observe a nucleon of
momentum $p_1$
\beq
\frac{(2\pi)^32p_{10}d\sigma}{d^3p_1}=\int dE'd\Omega
\sigma_{Mott}\frac{1}{4\pi M}
\left(\frac{M^2}{qP}F_2({\bf q,p_1})+2\tan^2\frac{\theta}{2}
F_1({\bf q,p_1})\right).
\eeq
Here the  structure functions $F_{1,2}({\bf q,p_1})$ are obtained from the
standard ones,
calculated in the preceding section, by lifting the integration over
$p_1$, that is
discarding $d^3p_1/((2\pi)^32p_{10}$ in the integration volume $dV$
in (34) and (38),
and also changing $p_{1y}$ and $p_{1z}$ to $\bar{p}_{1y}$ and
$\bar{p}_{1z}$  as
indicated in (46). Note that angles $\phi$ and $\phi_1$ in the latter
formulas depend
on the remaining integration variables, related to the changing vector
$\bf q=l-l'$.

Using (7) we rewrite (2) in terms of $W$'s to 
 obtain
\beq
\frac{(2\pi)^32p_{10}d\sigma}{d^3p_1}=\frac{1}{4\pi Am }\int dE'd\Omega
\sigma_{Mott}
\left(\frac{m^2}{p_+^2}\bar{W}_{++}(q,p_1))+2\tan^2\frac{\theta}{2}
\bar{W}_{yy}(q,p_1)\right). 
\eeq
In $W$'s we have to lift the integration over $\bf p_1$. We rewrite the phase
volume (26) in the form
\[
d\tau=\frac{d^3p_1}{(2\pi)^32p_{10}}\frac{d^4p_2}{(2\pi)^4}
2\pi\delta(p_2^2-m^2)
(2\pi)^4\delta^4(p_1+p_2-q-2p)
\]
and integrating over $p_2$ find
\beq
d\tau=\frac{d^3p_1}{(2\pi)^32p_{10}}
2\pi\delta((2p+q-p_1)^2-m^2),
\eeq
so that lifting the integration over $\bf p_1$ substitutes
\beq
d\tau\to 2\pi\delta((2p+q-p_1)^2-m^2).
\eeq

In variables $E',\theta,\phi$ we have
\[
q_0=E-E',\ \ q_z=-E+E'\cos\theta,\ \ q_x=-E'\sin\theta\cos\phi,\ \ 
q_y=-E'\sin\theta\sin\phi,\]
\beq
Q^2=2EE'(1-\cos\theta).
\eeq
The fixed momentum $\bf p_1$ can always be placed in the $xz$ plane,
so that $p_{1y}=0$. The argument of the $\delta$-function in (49) turns
out
to be
\beq
(2p+q-p_1)^2-m^2=4m^2-Q^2+4pq-4pp_1-2qp_1=
=2(b-E'\lambda(\theta,\phi)),
\eeq
where
\beq
\lambda=E(1-\cos\theta)+2m-E_1-p_{1z}\cos\theta+p_{1x}\sin\theta\cos\phi
\eeq
and
\beq
b=2m(m+E-E_1)-E(E_1+p_{1z}).
\eeq
Here  $E_1=\sqrt{m^2+p_{1z}^2+p_{1x}^2}$ is the energy of the observed
nucleon. Integrating this function we fix $E'(\theta,\phi)$ and
obtain an extra factor $1/(2\lambda)$.
Thus we get
\beq
\frac{(2\pi)^32p_{10}d\sigma}{d^3p_1}=\frac{1}{4 Am }\int 
\frac{d\cos\theta d\phi}{\lambda(\theta,\phi)} 
\sigma_{Mott}
\left(\frac{m^2}{p_+^2}\bar{W}_{++}(q,p_1))+2\tan^2\frac{\theta}{2}
\bar{W}_{yy}(q,p_1)\right), 
\eeq
where it is understood that we have to drop $d\tau$ in $W$'s, make the
substitutions (46) and express $E'$ in terms of $\theta$ and $\phi$
using (52):
\beq
E'=b/\lambda(\theta,\phi).
\eeq

Both $b$ and $\lambda(\theta,\phi)$ result positive and do not vanish in
the integration region. However $\sigma_{Mott}$ contains denominator $Q^4$
typical for virtual-photon-mediated processes and leading to a logarithmic
divergence for massless leptons $m_l=0$. To cure this divergence
we take in (55) the expression for $Q^2$ for non-zero $m_l$:
\beq
Q^2=2EE'-2\sqrt{(E^2-m_l^2)({E'}^2-m_l^2)}\cos\theta-2m_l^2.
\eeq

Note that
the standard
Weiszaecker-Williams
approximation takes into account only the leading term in $\log(1/m_l)$ and
thus expresses
the cross-section via the real photoproduction. However calculations
show that the
Weiszaecker-Williams contribution is generally quite a small part of
the total
cross-section due to the fact that the amplitude vanishes at $Q^2=0$.
This implies that leptoproduction cross-sections cannot be directly
reduced to photoproduction ones.

Lifting one or both integrations in (55) and introducing appropriate
phase volume factors one obtains cross-sections
with $Q^2$ or  4-vector $q$ fixed respectively, averaged over the azimuthal
directions of the observed nucleon. If one is interested also in this
azimuthal dependence, one should use the formula for the double
inclusive cross-section derived in Appendix 1.

\section{Photoproduction of cumulative protons and the choice of the
potentials}
Expressions for $H$ contain the relativistic potential $v(k)$
which desribes the pair interaction between the nucleons.
A variety of relativistic potentials have been proposed which
correctly describe the nucleon-nucleon scattering data at moderate
energies. However, on the one hand, we actually need the
high-energy asymptotics of the potential,
which is fixed by these data not well enough. On the other hand, the
proposed potentials refer to relativistic nucleons with spins
and it is not clear how they can be used in our spinless picture.
For this reason we shall recur to
a different approach. We shall use the fact that for the process of the
production
of  cumulative protons by real photons, the cross-sections are
directly related
to the internucleon potential. Therefore staying in the spinless
picture and using the experimental data on
the photoproduction
on deterium one can extract an effective internucleon potential 
as a function of the momentum transfer in a
straightforward manner. However the energy dependence of this
potential cannot be determined from the data, since the extracted
potential refers to the energies close to the threshold.
We studied two possibilities: an energy independent potential,
similar to the exchange of scalar mesons, and a potential which
grows with energy like the exchange of vectors mesons. Note that from
a purely theoretical point of view the first alternative seems 
preferable, since it leads to vanishing of finite state
interactions at high $Q^2$ in correspondence with colour transparency
following from the QCD. Also, as we shall see, our numerical results
with the energy independent potential agree quite satisfactorily with
the existing data on cumulative nucleon production, whereas with
the vector-like potential they overshoot the data by factor 2.5
(see Fig. 11 below). For these reasons, we discuss below only the
extraction from the photoproduction of an energy independent potential.

For unpolarized photons the total cross-section is
\beq
\sigma^{tot}=\frac{1}{J}e^2\sum_{pol}e^{\mu}e^{\nu}W_{\mu\nu},
\eeq
where $J=4Apq$ is the invariant flux, $e^{\mu}$ is the photon
polarization vector and $W_{\mu\nu}$ are the same as in Section 2.

We direct the momentum of the coming photon along the $z$ axis.
The integrand for $W_{\mu\nu}$ involves a product $H_{\mu}{H'_{\nu}}^*$ so 
that summing over polarization we get $(eH)(eH)^*=H(H')^*_{\perp}$.
From (18) we find that diagrams $c$ and $d$
do not contribute to the
cross-section in our approximation. Assuming that at $Q^2=0$ only the proton 
interacts with the incoming photon, we are left with  diagram $b$ only
(the direct mechanism) for which
\beq
H_{\perp}=2p_{1\perp}\frac{v(p_1-q-p)}{2qp_1}.
\eeq
This obviously implies that there is no rescattering and we can safely
drop small nuclear momenta in $H$. Repeating our derivation in Section 2 we find
the total cross-section as
\beq
\sigma^{tot}=\frac{1}{2}(A-1)\rho_0\frac{e^2}{8pq}
\int d\tau H^2_{\perp}.
\eeq

Passing to the inclusive cross-section for the production of a proton we 
make the substitution (50) to find
\beq
\frac{(2\pi)^32p_{10}d^3\sigma}{d^3p_1}= 
\frac{1}{2}(A-1)\rho_0\frac{e^2}{8pq}
2\pi\delta\Big((2p+q-p_1)^2-m^2\Big)H^2_{\perp}.
\eeq

We shall use the lab system where the nucleus is at rest. Then $qp=mE$ where
$E$ is the photon energy.
The argument of the $\delta$-function in (61) is  $ t+4mE-4mE_1+3m^2$,
where
\beq
 t=(q-p_1)^2=m^2-2E(E_1-p_1\cos\theta).
\eeq
 $E_1$ is the proton energy and $\theta$  the angle between the
proton and photon directions.

Transforming the phase volume element to variables $t,E_1$ and using the
$\delta$ function to integrate over $E_1$ we find
\beq
\frac{d\sigma}{dt}= \frac{1}{2}(A-1)\rho_0\frac{e^2}{256\pi m^2E^2}
H^2_{\perp}.
\eeq

Note that in $H_{\perp}$ the argument of the potential is
\[
 (p_1-q-p)^2=t+m^2+2m(E-E_1)=(1/2)(t-m^2)\equiv(1/2)t_1
\]
and the denominator is
$2qp_1=-t_1$, so that we finally obtain
\beq
\frac{d\sigma}{dt}= \frac{1}{2}(A-1)\rho_0
\frac{e^2p_{1\perp}^2}{64\pi m^2E^2}
\left(\frac {v(t_1/2)}{t_1}\right)^2.
\eeq

For the deuteron we have to substitute
\beq
\frac{1}{2}A(A-1)\rho_0=(1/m)\psi^2(0),
\eeq
where $\psi(r)$ is the deuteron wave function in the coordinate space.
 
As we observe, the photoproduction cross-section reveals a great
simplicity in
our approach: due to kinematical reasons the rescattering is absent and the
cross-section becomes directly related to the high-momentum asymptotics 
of the effective 
internucleon potential. Therefore it is most suitable for studying this 
asymptotics.
Of course one has to remember that our formulas may pretend to be valid 
only for 
the values of $x$ sufficiently higher than unity. This is achieved in 
the emission
of protons in the backward hemisphere relative to the direction of 
the incoming photon.
Present data mostly refer to forward directions. Only the data [ 9 ] on the
deuteron at  the incident photon energy 4.0 GeV and
$\theta_1=90^o$ 
involve  sufficiently high values of $x$. These data show 
that with more or 
less standard nuclear potentials borrowed from low energy physics one 
cannot desribe the  experiment
above $E=1$ GeV.
The experimental data for $s^{11}d\sigma/dt$ 
show a clear plato above
1.4 GeV in accordance with
the QCD scaling law, whereas the predictions from the meson-exchange
potentals steadily rise [ 9 ]. This is not at all unexpected, since one
cannot hope that the relatively low energy parametrizations for the
potential will work sufficiently well at high momentum transfers
involved.

As mentioned, we are going to use the relation between the photoproduction 
cross-section and potential
in the opposite direction. We determine the asymtotics of the effective
nuclear potential from Eq. (64) putting the experimental data from [ 9 ]
into its left-hand side.
The latter can be satisfactorily fit by the expression
\beq
s^{11}\frac{d\sigma}{dt}\, (kbn*GeV^{20})=0.4+0.147741
e^{-E^2}+1.47116e^{-2E^2}
\eeq 
valid for $E=0.2 \div 4.0$ GeV.
We take the value for the deuteron wave function at the origin as
\[ \psi(0)=3.131\, 10^{-2} {\rm GeV}^{3/2}.\]
The momentum space potential extracted in this manner is shown in Fig. 3. 
Compared to standard internucleon potentials it
falls more rapidly with the transferred momentum. 
This is in agreement with the reduced nuclear amplitude approach of [ 10 ], 
in which hard exchanges between nucleons are to be accompanied by
damping form-factors.

\section{Numerical results}
Using the effective potential extracted from the photoproduction data
we calculated the leptoproduction cross-sections (55) for the
initial lepton energy 6 GeV in the backward hemisphere on various nuclear
targets. The $A$ dependence of our cross-sections is concentrated in
factors $\rho_0$ and $l$ for short- and long-range contributions
correspondingly. These factors were calculated in the standard manner,
neglecting correlations in the nuclear $\rho$-matrix and using
the Woods-Saxon nuclear density. One finds
\beq
(A-1)\rho_0=\frac{1}{2m}A(A-1)\int d^3r\rho^2_{NR}(r)\equiv m^2f_0,
\eeq
\beq
\frac{2}{A}l=\frac{1}{2m}A(A-1)\int d^2bT^2(b)\equiv mc_0.
\eeq
Calculated dimensionless $f_0$ and $c_0$  are shown in Fig. 4
for different nuclei.

Numerical integrations in (55) encounter certain difficulties related to the
singularities of the integrand at $Q^2=0$ and the values of angle $\theta$
limiting the rescattering region. Some details on this point are given
in the Appendix 2.

Our results for the inclusive cross-section are presented in Figs. 5
and 6 for Cu and Pb targets for various angles in the backward hemisphere.

To see the relative importance of different production mechanisms,
in Figs. 7-10 we show
differential cross-sections in both ${\bf p}_1$
and $Q^2$ at $Q^2=1$ and 4 (GeV/c)$^2$ and 90$^o$ and 180$^o$ on
the Pb target. Solid curves correspond to the total
contribution. Dashed and dash-dotted curves show contributions from
the spectator mechanism and rescattering correspondingly.

As a general result we find that for the studied values of $Q^2$
in the kinematical region where rescattering is possible it either
completely dominates (at $Q^2\sim 1$ (GeV/c)$^2$ or lower) or
gives a contribution of the same order as the spectator mechanism.
The contribution of the direct mechanism has been always
found  quite small. At comparatively low values of $Q^2$  the short
range contribution from the final interaction ("compressed tube"
mechanism) is also of the same order as the spectator mechamism
and negative (see Fig. 8). However with the growth of $Q^2$ its
relative weight diminishes, as expected (cf. Fig. 10).
We have also found that the relative contribution
from the Weizsaecker-Williams term
(leading  in $\log (1/m_l)$) is generally quite small, except at 
highest values of $p_1$ for a given angle where it becomes of the same order
or even several times larger than the rest short-range contribution.

To have some contact with existing experimental data on A(e,e'p) reaction
we also calculated the double inclusive cross section
 $d\sigma/(dE'd\Omega d\Omega_1)$
on the $^{16}$O target,
where $\Omega$ and $\Omega_1$ are angular variables for the final lepton
and nucleon respectively. We have taken $E=2.4$ GeV, $Q^2=0.8$ (GeV/c)$^2$
and $q_0=0.439$ GeV corresponding to the experimental kinematics in [ 11 ].
Our results for the backward hemisphere are shown in Fig. 11 together with
experimental points for the protons knocked out of 1p$_{1/2}$ shell
(lower points) and 1p$_{3/2}$ shell (upper points). The recoil nucleon
momenta result  small in this kinematics, so that our approximation of
neglecting the binding does not seem to be well justified.
However the agreement is unexpectedly good: our calculated cross-sections
well correspond to the experimental  ones averaged over
the two nuclear levels.

\section{Conclusions}
We have studied leptoproduction of cumulative nucleons on nuclear targets
in a realistic subasymptotic kinematics, where all proposed mechanisms
work together and interfere. Our results show that in the kinematical
region where rescattering is possible, its contribution cannot
be neglected unless $Q^2$ is fixed and large. For the inclusive cross-section
intergrated over $Q^2$ rescattering  dominates the cross-section in this
region. Of the rest mechanisms the direct one has been found to be
completely unimportant at $Q^2>1$ (GeV/c)$^2$. However at smaller $Q^2$
its contribution interferes with the rest mechanisms to make the amplitude
vanish at $Q^2=0$. The short range part of the final
state interaction (the compressed tube mechanism) gives a negative
contribution, of the same order that the spectator contribution, except
at high values of $Q^2$ when it diminishes.

So, on the one hand, our results confirm the standard expectations
that at very high $Q^2$ only the spectator mechanism survives, which
allows to simply relate the cumulative production on different
nuclear targets [ 8 ]. On the other hand, they show that for realistic
not too large $Q^2$ and certainly for the inclusive cross-sections
integrated over all $Q^2$ all mechanisms give contributions of
comparable order and interfere. In particular, in the kinematical region
accessible for rescattering the latter gives a sizable (or even dominating)
 part of the contribution,
which prevents using simple scaling arguments to fix the $A$-dependence.

Another important result is that the Weizsaecker-Williams
approximation generally constitutes only a small part of the
total cross-section.
Thus leptoproduction cross-sections cannot be simply reduced to
photoproduction ones.

Our calculations have been based on a simplified picture,
in which  all spins have been neglected. Accordingly the effective
internucleon relativistic potential  have been taken
from the experimental photoproduction cross-sections, calculated
within the same spinless approach. Comparison with the existing data
on A(e,e'p) reaction 
reveals a reasonable agreement, which supports validity of our treatment.
\section{Acknowledgements}

M.A.B. is deeply thankful to the NCCU NC for hospitality and attention.
This study was partially supported by the grant RFFI (Russia) 01-02-17137

\section{Appendix 1. The double inclusive cross-section}
In principle lifting integrations in the inclusive cross-sections
does not give the exclusive ones: one can add any function to the
thus obtained cross-section which integrates out to zero in the inclusive
cross-section. So to have the double cross-section for the process
in whichboth the
lepton and nucleon are observed we have
to study the amplitudes as they are, not via the inclusive form
with the structure functions $F_{1,2}$. In our case this is possible
(since the amplitudes are known).

The starting point is the basic formula for the cross-section
\beq
d\sigma=\frac{1}{4(Pl)}\frac{e^4}{q^4}L^{\mu\nu}W_{\mu\nu}d\tau
\eeq
in which $4Pl$ is the flux, $L$ and $W$ are the leptonic and hadronic tensors
and $d\tau$ is the phase volume.
The leptonic tensor is
\beq
L_{\mu\nu}= \frac{1}{2}\sum_{pol}(\bar{u}'\gamma_{\mu}u)
(\bar{u}\gamma_{\nu}u')=2(l'_{\mu}l_{\nu}+l_{\mu}l'_{\nu}-(ll')g_{\mu\nu}).
\eeq
The hadronic tensor can be represented by a symbolic product of
two currents $J$ which are in fact matrix elements of electronagnetic
current between the initial and final hadronic states:
\beq
W_{\mu\nu}=J_{\mu}J^*_{\nu}.
\eeq
Thus the cross-sections has the form
\beq
d\sigma=\frac{1}{2(Pl)}\frac{e^4}{q^4}d\tau
\Big[ (l'J)(lJ)^*+(lJ)(l'J)^*-(ll')(JJ^*)\Big].
\eeq

We choose the system $l_+=l_{\perp}=0$. Then both the numerator and
denominator become proportional to $l_-$, which cancels out. We get
\[
d\sigma=\frac{1}{2P_+}\frac{e^4}{q^4}d\tau
\Big[ (l'J)J_+^*+J_+(l'J)^*-l'_+(JJ^*)\Big]\]\beq
=\frac{1}{2P_+}\frac{e^4}{q^4}d\tau
\Big[ (2l'_-J_+J_+^*+l'_+(JJ^*)_{\perp}-(l'J)_{\perp}J^*_+-
J_+(l'J)_{\perp}^*\Big].
\eeq
Now we use that
\beq
l'_+=\frac{1}{\sqrt{2}}E'(1-\cos\theta)=\sqrt{2}E'\sin^2\frac{\theta}{2},
\ \
l'_-=\frac{1}{\sqrt{2}}E'(1+\cos\theta)=\sqrt{2}E'\cos^2\frac{\theta}{2}.
\ \
\eeq
We also direct $q_{T}$ along the $x$-axis, so that
\beq
l'_x= -E'\sin\theta.
\eeq
The cross-section becomes
\[
d\sigma=\frac{1}{2P_+}\frac{e^4}{q^4}\sqrt{2}E'\cos^2\frac{\theta}{2}d\tau
\Big[ (2J_+J_+^*+\tan^2\frac{\theta}{2}(JJ^*)_{\perp}+
\sqrt{2}\tan\frac{\theta}{2}(J_xJ^*_++J_+J^*_x)\Big]\]\beq=
\frac{1}{2P_+}\frac{e^4}{q^4}\sqrt{2}E'\cos^2\frac{\theta}{2}d\tau
\Big[ (|\sqrt{2}J_++\tan\frac{\theta}{2}J_x|^2
+\tan^2\frac{\theta}{2}|J_y|^2\Big].
\eeq
The first factor in (76) is
\beq 
\frac{1}{2P_+}\frac{e^4}{q^4}\sqrt{2}E'\cos^2\frac{\theta}{2}=
\frac{4\pi^2}{ME'}\sigma_{Mott}.
\eeq
We present
\beq
d\tau=\frac{d^3l'}{(2\pi)^32E'}d\tau_h,
\eeq
where $d\tau_h$ is the phase volume of the observed  hadron.
Then we finally get the exclusive cross-section as
\beq
d\sigma=dE'd\Omega'd\tau_h
\frac{\sigma_{Mott}}{4\pi M}
\Big[ (|\sqrt{2}J_++\tan\frac{\theta}{2}J_x|^2
+\tan^2\frac{\theta}{2}|J_y|^2\Big].
\eeq

To compare with the cross-section which is found by lifting the
integrations in the inclusive cross-section we transform (79) to the system
$q_T=0$. This is achieved by a rotation in $xz$ plane by angle $\theta_0$
determined by
\beq
\tan\theta_0=-\frac{q_x}{q_z}=\frac{E'\sin\theta}{E-E'\cos\theta}.
\eeq
We find
\beq
d\sigma=dE'd\Omega'd\tau_h\frac{\sigma_{Mott}}{4\pi M}
\Big[ (|Z|^2+\tan^2\frac{\theta}{2}|J_y|^2\Big],
\eeq
where now
\beq
Z=c_+J_++c_-J_-+c_xJ_x
\eeq
with
\beq
c_{\pm}=\frac{1}{\sqrt{2}}\left(1\pm\cos\theta_0\mp
\sin\theta_0\tan\frac{\theta}{2}\right),\ \
c_x=\sin\theta_0+\cos\theta_0\tan\frac{\theta}{2}.
\eeq

On the other hand, starting from the inclusive cross-section (2),
dropping the integration over $d\tau_h$ and passing from the system $q_+=0$
to the system $q_T=0$ we obtain
\beq
d\tilde{\sigma}=dE'd\Omega'd\tau_h\frac{\sigma_{Mott}}{4\pi M}
\Big[ (|\tilde{Z}|^2+2\tan^2\frac{\theta}{2}|J_y|^2\Big]
\eeq
with
\beq
\tilde{Z}=\tilde{c}_+J_++\tilde{c}_-J_-+\tilde{c}_xJ_x,
\eeq
where
\beq
\tilde{c}_{\pm}=\frac{1}{\sqrt{2}}(1\pm\cos\phi_0),
\ \
\tilde{c}_x=\sin\phi_0
\eeq
and
\beq
\tan\phi_0=\frac{xM}{Q}.
\eeq

One can trivially prove that
\beq
\cos\phi_0=\cos\theta_0-\sin\theta_0\tan\frac{\theta}{2},\ \
\sin^2\phi_0+\tan^2\frac{\theta}{2}=\left(\sin\theta_0+
\cos\theta_0\tan\frac{\theta}{2}\right)^2.
\eeq
So it turns out that
\beq
\tilde{c}_{\pm}=c_{\pm},\ \
\tilde{c}_x=\sqrt{c^2_x-\tan^2\frac{\theta}{2}}.
\eeq

Comparing the correct cross-section (81) with the one (84) found from the
inclusive cross-section, we see that they do not coincide, due to the
difference in $c_x$ and $\tilde{c}_x$. However if we integrate
over the azimuthal directions of the observed nucleon, then
both $|J_x|^2$ and $|J_y|^2$ go over to $(1/2)|J_\perp|^2$ and the
interference term with $J_+J_x^*$ vanishes. Then both cross-sections
(81) and (84) give the same result.

Thus for the cross-sections averaged over the azimuthal directions
of the observed nucleon one can use the inclusive cross-section
with the dropped intergration over $p_{1z}$ and $p_{1\perp}$.
However if one wants to study also the azimuthal dependence, then
one has to use the correct expression (81).
\section{Appendix 2. Some calculational details}

\subsection{Singularity at $\theta=0$}
We present the cross-section (55) in the form 
\beq
\frac{(2\pi)^32p_{10}d\sigma}{d^3p_1}=\frac{\alpha_{em}^2}
{4 AmE^2 }\int 
\frac{d\cos\theta d\phi}{(1-\cos\theta)^2}X(\theta,\phi),
\eeq
where
\beq
X(\theta,\phi)=\frac{\cos^2(\theta/2)}
{\lambda(\theta,\phi)} 
\left(2\bar{W}_{++}(q,p_1)+\tan^2\frac{\theta}{2}
\bar{W}_{yy}(q,p_1)\right). 
\eeq
With a zero lepton mass, the integral (90) diverges
logarithmically at $\theta=0$ (corresponding to $Q^2=0$, a real
photon). With a non-zero lepton mass $m_l$ the divergence disappears,
but the integrand still results strongly peaked at $\theta=0$.
To make its numerical calculation feasible one has naturally to
separate the peak from the background.

Presenting at small $\theta$
\beq
X(\theta,\phi)=\frac{1}{2}\theta^2X_1(\phi)=(1-\cos\theta)X_1(\phi)
+\left(\cos\theta-1+\frac{1}{2}\theta^2\right)X_1(\phi),
\eeq
we write the integral (90) in the form
\beq
\frac{(2\pi)^32p_{10}d\sigma}{d^3p_1}=\frac{\alpha_{em}^2}
{4 AmE^2 }\int 
d\cos\theta d\phi\Big[\frac{(1-\cos\theta)X_1(\phi)}
{(1-\cos\theta+\mu^2)^2}+
\frac{X(\theta,\phi)-(1-\cos\theta)X_1(\phi)}{(1-\cos\theta)^2}
\Big],
\eeq
where 
\[\mu=\frac{m_l}{\sqrt{2}}\left(\frac{1}{E}+\frac{1}{E'}\right).\].

In the first term 
the integrations over $\theta$ and $\phi$ can be done explicitly.
The result corresponds to the Weiszsaecker-Williams approximation.
From our study it follows that the dependence of $X_1(\phi)$
on the azimuthal angle is trivial
\beq
X_1(\phi)=\cos^2\phi X_{10}^{(1)}+\sin^2\phi X_{10}^{(2)},
\eeq
where terms with $cos$ and $sin$ come from $W_{++}$ and
$W_{yy}$ respectively.
So the integration over $\phi$ leads to the substitution
\beq
X_1(\phi)=\to \pi (X_{10}^{(1)}+X_{10}^{(2)})=\pi X_{10}
\eeq
Integrating over $\cos\theta$ we finally get
\beq
\frac{(2\pi)^32p_{10}d\sigma}{d^3p_1}=
\frac{(2\pi)^32p_{10}d\sigma^{WW}}{d^3p_1}+
\frac{\alpha_{em}^2}
{4 AmE^2 }\int 
d\cos\theta d\phi
\frac{X(\theta,\phi)-(1-\cos\theta)X_1(\phi)}{(1-\cos\theta)^2},
\eeq
where the Weizsaecker-Williams term is given by
\beq
\frac{(2\pi)^32p_{10}d\sigma^{WW}}{d^3p_1}=
\frac{\pi\alpha_{em}^2}
{4 AmE^2 } X_{10}\left(\ln\frac{2}{\mu^2}-1\right).
\eeq
The integration over $\theta$ and $\phi$ in the second term in (96)
can only be done numerically. 

\subsection{Singularities of the short-range part of $X$}
In the rescattering region the calculation of the short-range 
contribution to the integral (96)
encounters difficulties generated by the singularities of 
Re $H_+^2$ at $(q+p)^2=m^2$. 
 Denote 
\[ X_r(\theta,\phi)=X(\theta,\phi)-(1-\cos\theta)X_1(\phi).\]
As a function of $\phi$ $X_r$ has poles of the 1st and 2nd order
in $\cos\phi$:
\beq
X_r(\theta,\phi)=X_{rr}(\theta\phi)
+\frac{r_1(\theta)}{A(\theta)-B(\theta)\cos\phi}+
\frac{r_2(\theta)}{\Big(A(\theta)-B(\theta)\cos\phi\Big)^2},
\eeq
where $X_{rr}$ is a regular function. The poles
exist only in a certain interval of $\theta$
\beq
\theta_{min}<\theta<\theta_{max}
\eeq
determined by the condition 
\beq
\left|\frac{A(\theta)}{B(\theta)}\right|<1,
\eeq
where rescattering is kinematically possible.

Integration of  Re $X_r$ over $\phi$ evidently gives
\beq
\int d\phi {\rm Re}\, X_r(\theta,\phi)=
\int d\phi{\rm Re}\,X_{rr}(\theta,\phi),
\eeq
for $\theta$ inside the rescattering interval (99),
and
\beq
\int d\phi {\rm Re}\, X_r(\theta,\phi)=
\int d\phi{\rm Re}\, X_{rr}(\theta,\phi)
+2\pi\left(\frac{1}{\sqrt{A(\theta)^2-B(\theta)^2}}+
\frac{B(\theta)}{[A(\theta)^2-B(\theta)^2]^{3/2}}\right)
\eeq
for $\theta$ outside the rescattering interval.
As we observe in the latter case the result is a 
non-integrable function of $\theta$ at points $\theta_{min}$ and
$\theta_{max}$.

To overcome this difficulty we have to remember that in reality
the double pole at $A-B\cos\phi$ is split into two close poles
when one takes into account the small nuclear momenta neglected in
in (102). Thus in the vicinity of, say, $\theta_{min}$ the integrand
in (96)
\beq
Y(\theta)=\frac{1}{(1-\cos\theta)^2}\int d\phi {\rm Re}, X_r(\theta\phi)
\eeq
will have a singularity structure
\beq
Y(\theta)=Y_r(\theta)
+\frac{1}{\alpha}\left(\frac{c}{\sqrt{\theta_{min}-\theta}}-
\frac{c+c_1\alpha}{\sqrt{\theta_{min}+t\alpha-\theta}}\right),
\eeq
with $Y_r$ a regular function and
$\alpha<<1$  proportional to the difference
$\bf (q,k_1-k'_1)$. In the limit $ \alpha\to 0$ we obtain from (104)
\beq
Y(\theta)_{\alpha\to 0}=Y_r(\theta) +\frac{\eta}{\sqrt{\theta_{min}-
\theta}}+
\frac{\zeta}{(\theta_{min}-\theta)^{3/2}},
\eeq
where
\[\eta=a-c_1,\ \ \zeta=ct/2,\]
with a non-integrable singularity. However, correctly integrated, (104)
has a finite limit at $\alpha\to 0$ due to the dependence on $\alpha$
of the limits of integration.
Indeed one trivially finds  in this limit
\beq
\int_{\theta_0}^{\theta_{min}}Y(\theta)=
\int_{\theta_0}^{\theta_{min}}Y_r(\theta)+2\eta
\sqrt{\theta_{min}-\theta_0}-\frac{2\zeta}
{\sqrt{\theta_{min}-\theta_0}}.
\eeq
Quite a similar formula is obtained for the integral over the
interval $\theta_{max}<\theta<\theta_1$ with the substitution
$\theta_{min}-\theta_0\to \theta_1-\theta_{max}$ and of course
different values for $\eta$ and $\zeta$.

In principle the regularization functions $r_{1,2}(\theta)$ and
numbers $\eta$ and $\zeta$ for the two singular points in $\theta$ can
be obtained analytically. However the resulting expressions are
very cumbersome, so that in practice we obtained them by 
studying numerically the corresponding singular behaviour.

An additional difficulty consists in that the lower singular 
point $\theta=\theta_{min}$ for values of $x_1$ close to unity
turns out to be quite small and so very close to the the point
$\theta=0$ at which function $Y_r$ aquires a singularity due
to subtraction. To solve this problem we divide the interval
$[0,\theta_{min}]$ into two $[0,\theta_0]+[\theta_0,\theta_{min}]$
and apply the regularization (106) only in the 2nd interval.
As a result  the integrations over these two
intervals  give large numbers  cancelling each other, which 
imposes very stringent requirements on the numerical precision.

\section{References}
1. D.I.Blokhintzev, JETP {\bf 33} (1957) 1295.

2. I.A.Scmidt and R.Blankenbecler, Phys. Rev.{\bf D 15} (1977) 3321;
 {\bf D 16} (1977) 1318

3. L.L.Frankfurt and M.I.Strikman, Phys. Rep. {\bf 76} (1981) 215

4. A.V.Efremov, Yad. Fiz. {\bf 24} (1976) 1208;\\
V.V.Burov, V.K.Lukianov and A.I.Titov, Phys. Lett. {\bf B67} (1977) 46;\\
A.V.Efremov, A.B.Kaidalov, V.T.Kim, G.I.Lykasov and N.V.Slavin,
Yad. Fiz. {\bf 47} (1988) 1364.

5.M.I.Gorenshtein, G.M.Zinoviev and V.P.Shelest,
Yad. Fiz. {\bf 26} (1977) 788.

6. L.A.Kondratiuk and V.B.Kopeliovich, JETP Lett. {\bf 21} (1975) 40;\\
V.B.Kopeliovich, JETP Lett. {\bf 23} (1976) 313, Phys. Rep. {\bf 139}
(1986) 51;\\
M.A.Braun and V.V.Vechernin, Sov. J. Nucl. Phys. {\bf 25} (1977) 676,
{\bf 28} (1978) 753.

7. M.A.Braun and V.V.Vechernin, J.Phys. {\bf G 19} (1993) 517, 531

8. L.L.Frankfurt, M.I.Strikman, D.B.day and M.Sargsyan, Phys. rev. {\bf C 48}
(1993) 2451. 

9. C.Bochna {\it et al.}, Phys. Rev. Lett. {\bf 81} (1998) 4576.

10. S.J.Brodsky and J.R.Hiller, Phys. Rev. {\bf C 28} (1983) 475.

11. J.Gao {\it et al.}, Phys. Rev. Lett. {\bf 84} (2000) 3265.

\section{Figure captions}

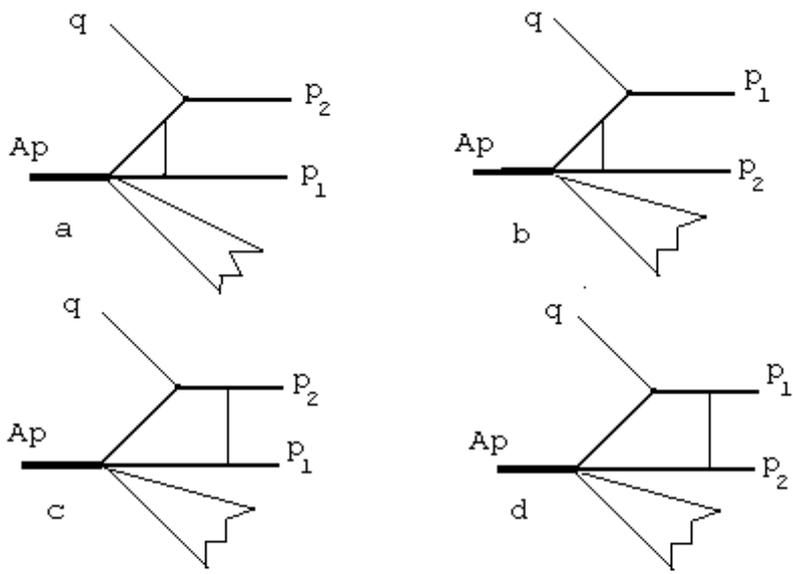
Fig. 1.
Diagrams illustrating different mechanisms of cumulative production:
spetator ($a$), direct ($b$), compressed tube plus rescattering
($c$ and $d$).

Fig. 2. Cumulative (solid line) and rescattering (dashed line) kinematical
regions in the backward hemisphere at 6 GeV.

Fig. 3. Effective internucleon momentum space potential as extracted
from the photoproduction data [ 9 ].

Fig. 4  Nuclear coefficients $f_0$ and $c_0$.

Fig. 5. Inclusive nucleon leptoproduction cross-sections on Cu
at the incident energy 6 GeV in the backward hemisphere.

Fig. 6. Same as Fig. 5 on the Pb target.

Fig. 7. Double differential cross-sections in ${\bf p}_1$ and $Q^2$
for nucleon leptoproduction on Pb at the incident energy 6 GeV,
emission angle $\theta_1=90^o$ and $Q^2=1$ (GeV/c)$^2$.

Fig. 8. Same as Fig. 8 at  $\theta_1=180^o$ and $Q^2=1$ (GeV/c)$^2$.

Fig. 9. Same as Fig. 8 at  $\theta_1=90^o$ and $Q^2=4$ (GeV/c)$^2$.

Fig. 10. Same as Fig. 8 at  $\theta_1=180^o$ and $Q^2=4$ (GeV/c)$^2$.

Fig. 11. Double inclusive cross-section for the reaction
$e+^{16}O\to e'(p_e)+N(p_1)+X$ at $E=2.4$ GeV, $Q^2=0.8$ (GeV/c)$^2$ and
$q_0=0.439$ GeV for the nucleons ejected into the backward hemisphere.
Positive $p_{miss}=|{\bf q-p_1}|$ correspond to $\phi_1=180^o$, negative
$p_{miss}$ to $\phi_1=0$. The lower curve was obtained with an
energy independent internucleon potential, the upper curve
with a vector-like potential. Experimental
points are from [ 11 ].
Lower (upper) points correspond to the 1p$_{1/2}$ (1p$_{3/2}$) nuclear
level.

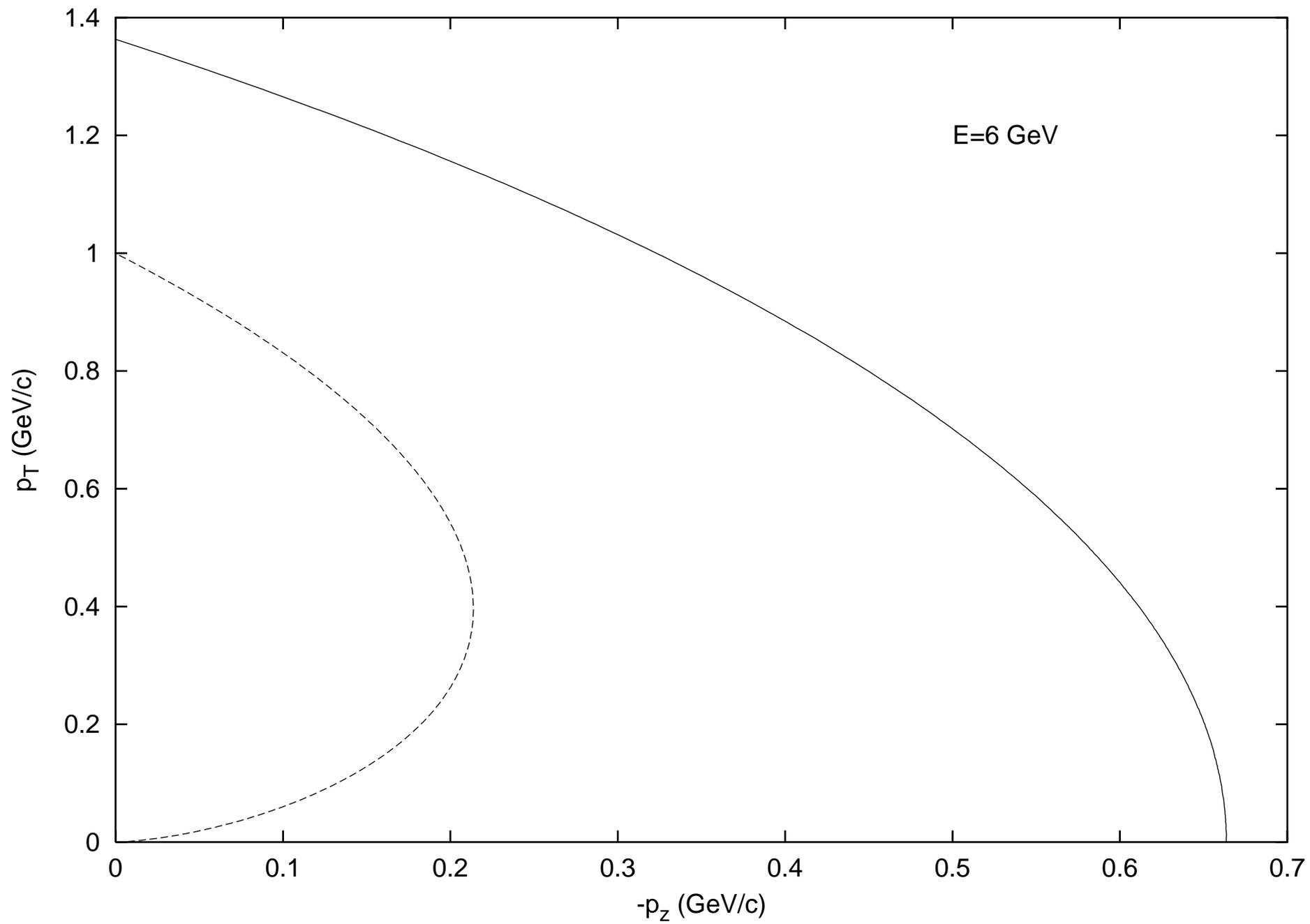
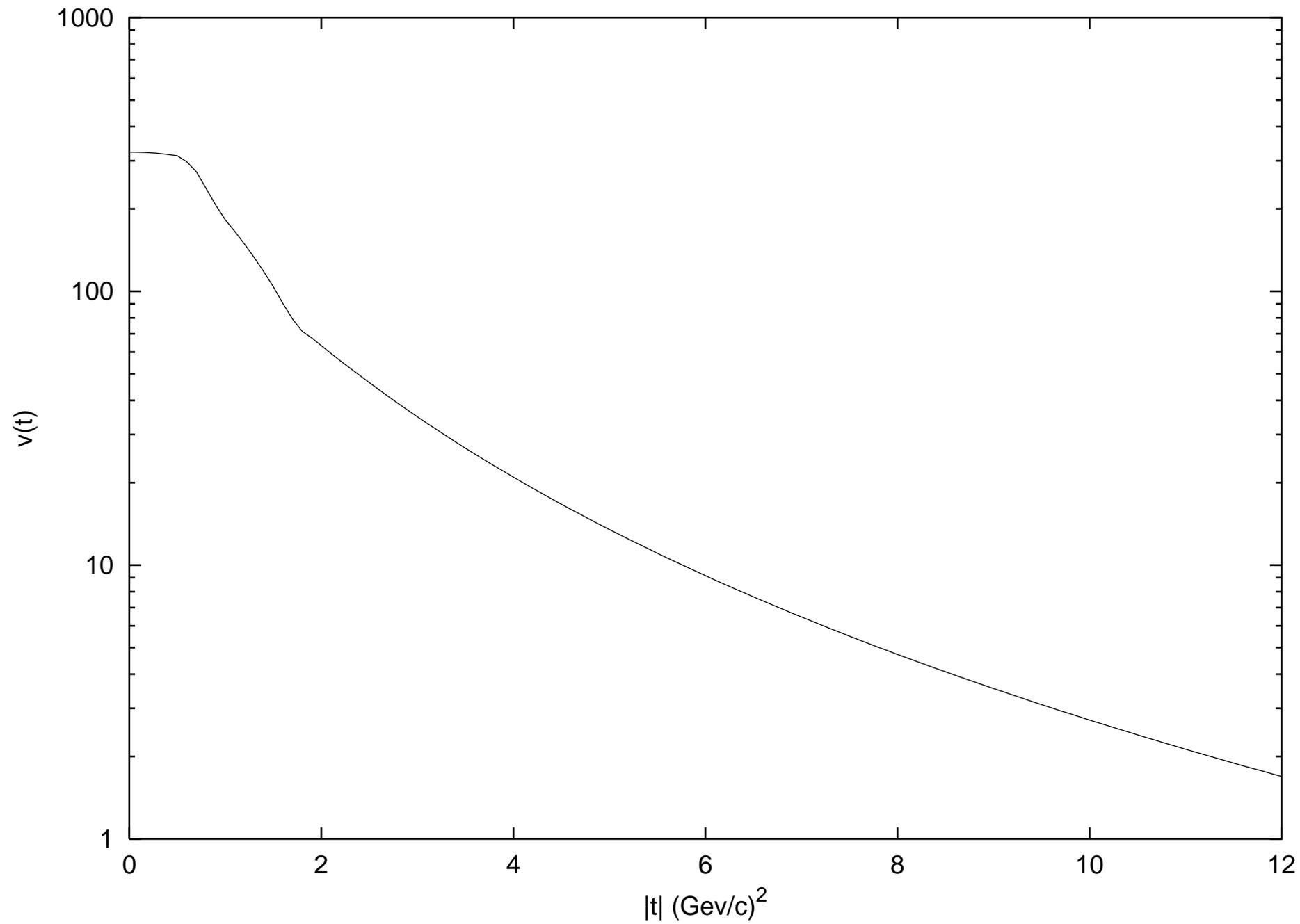
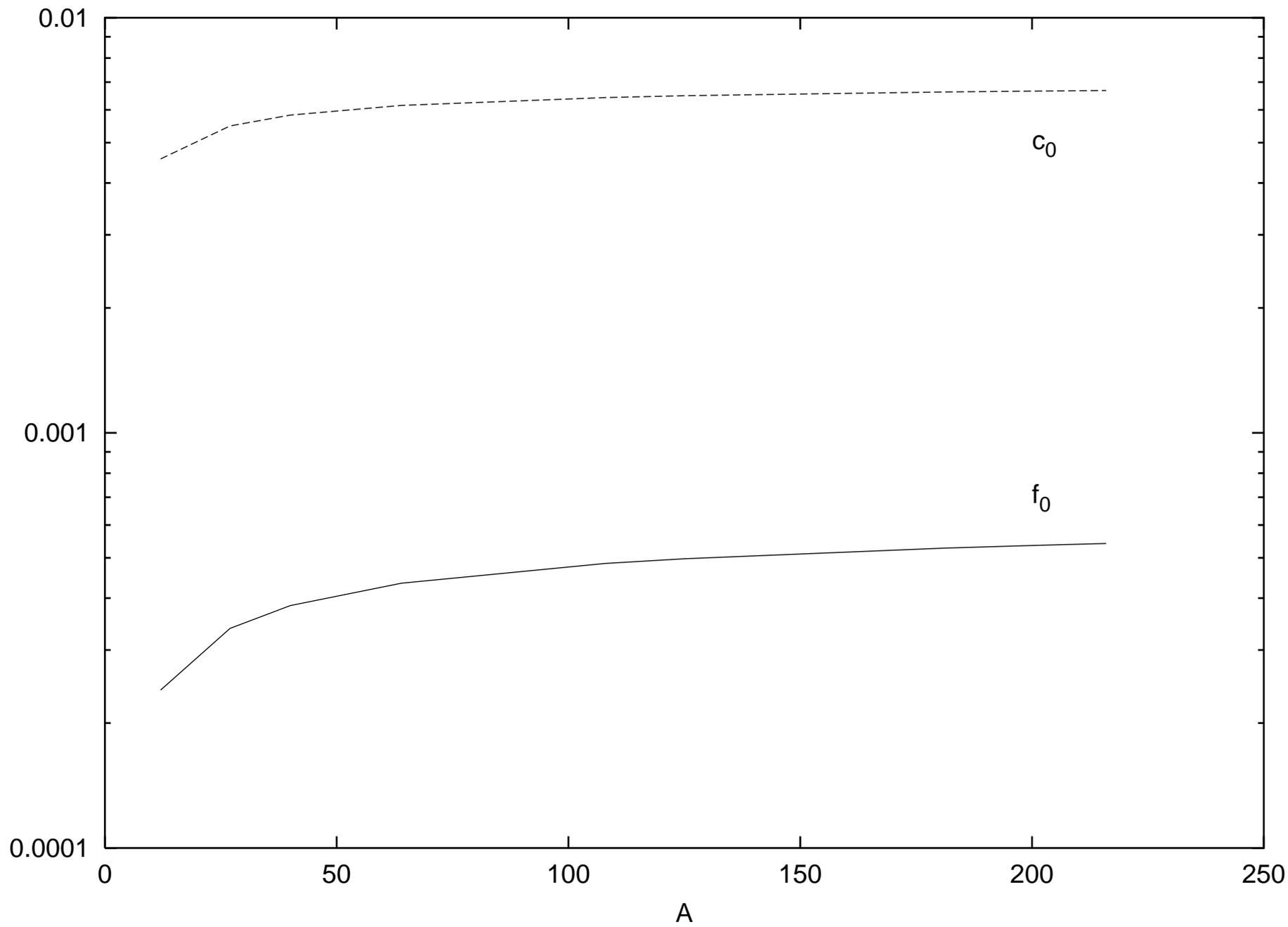
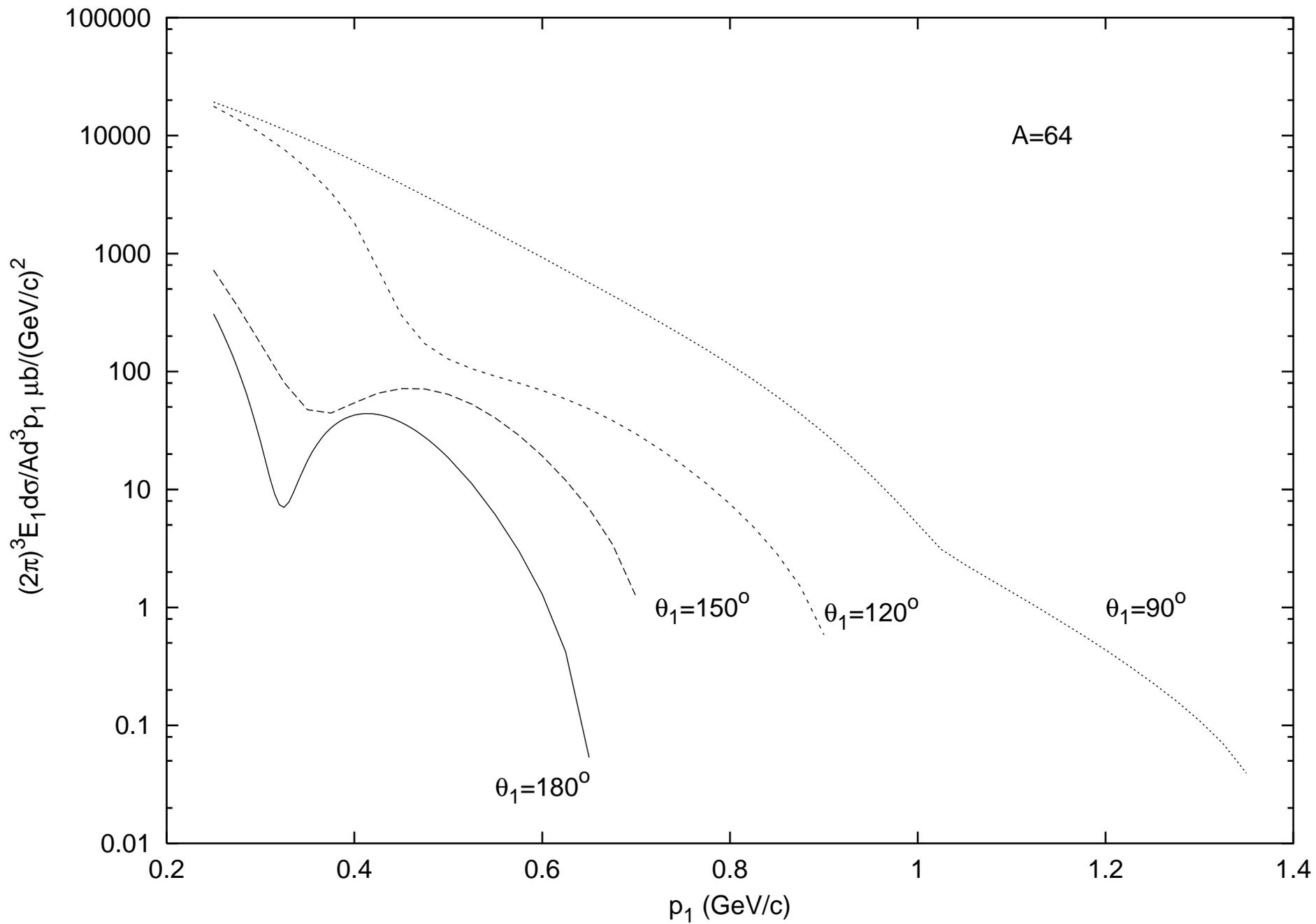
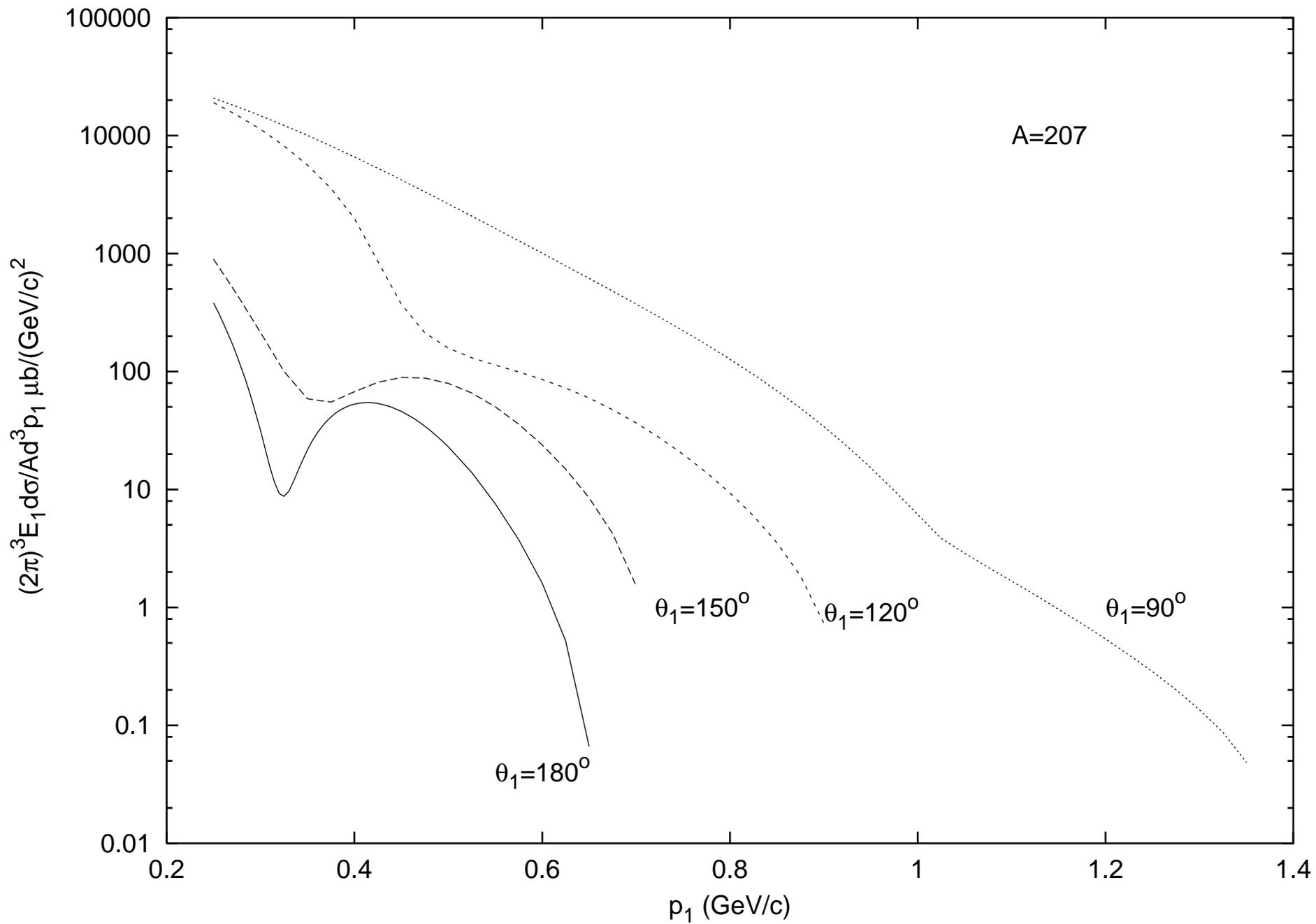
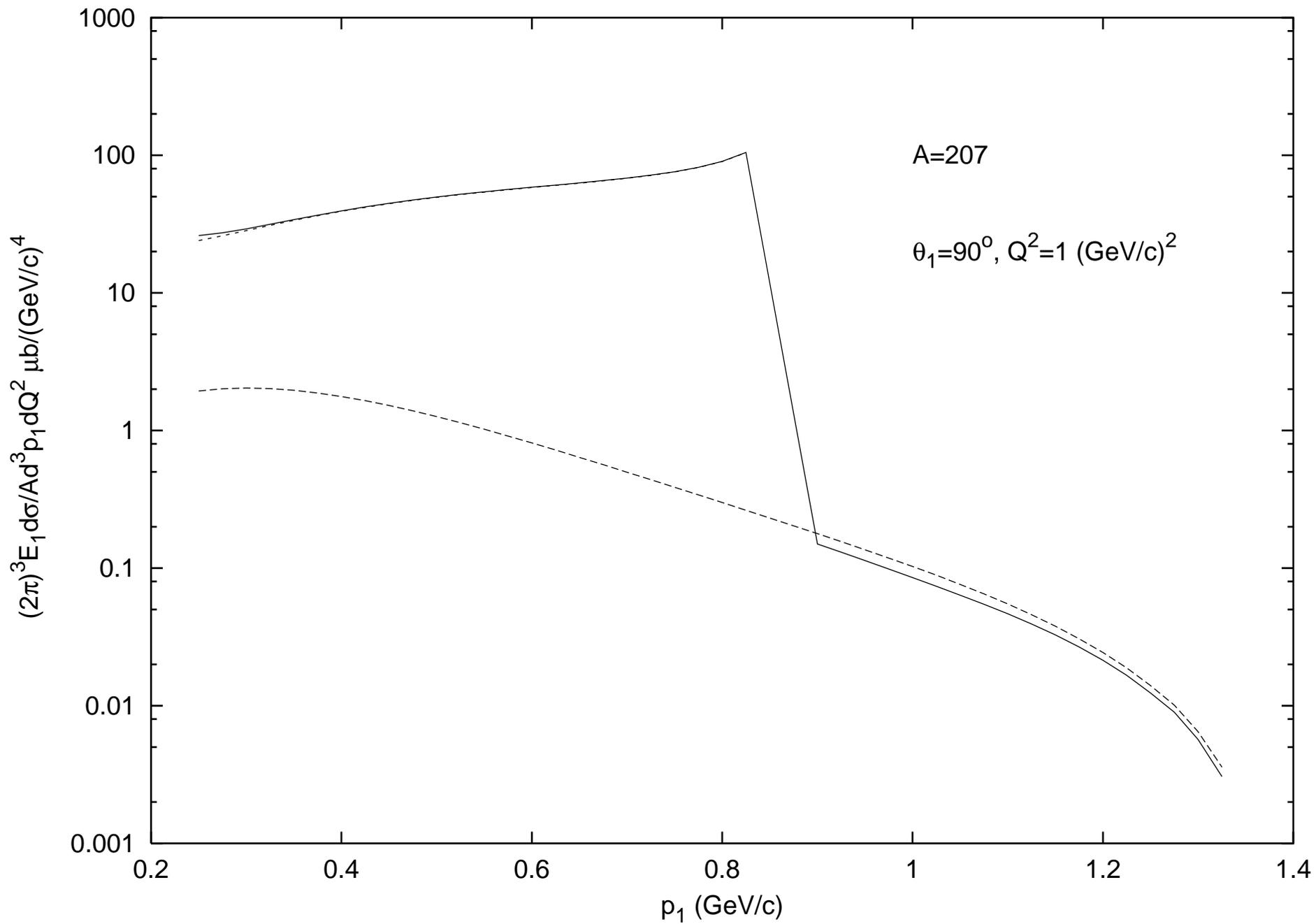
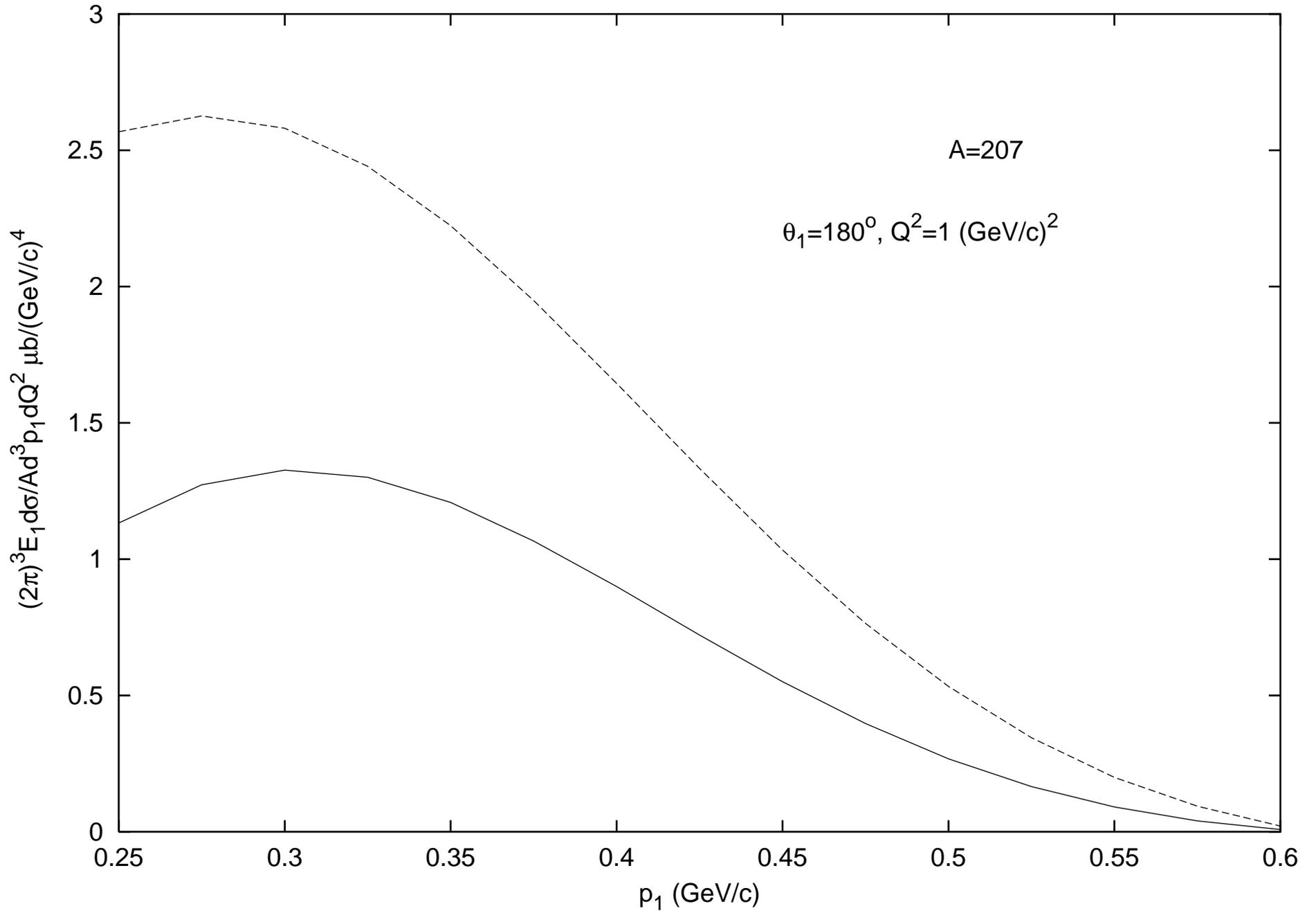
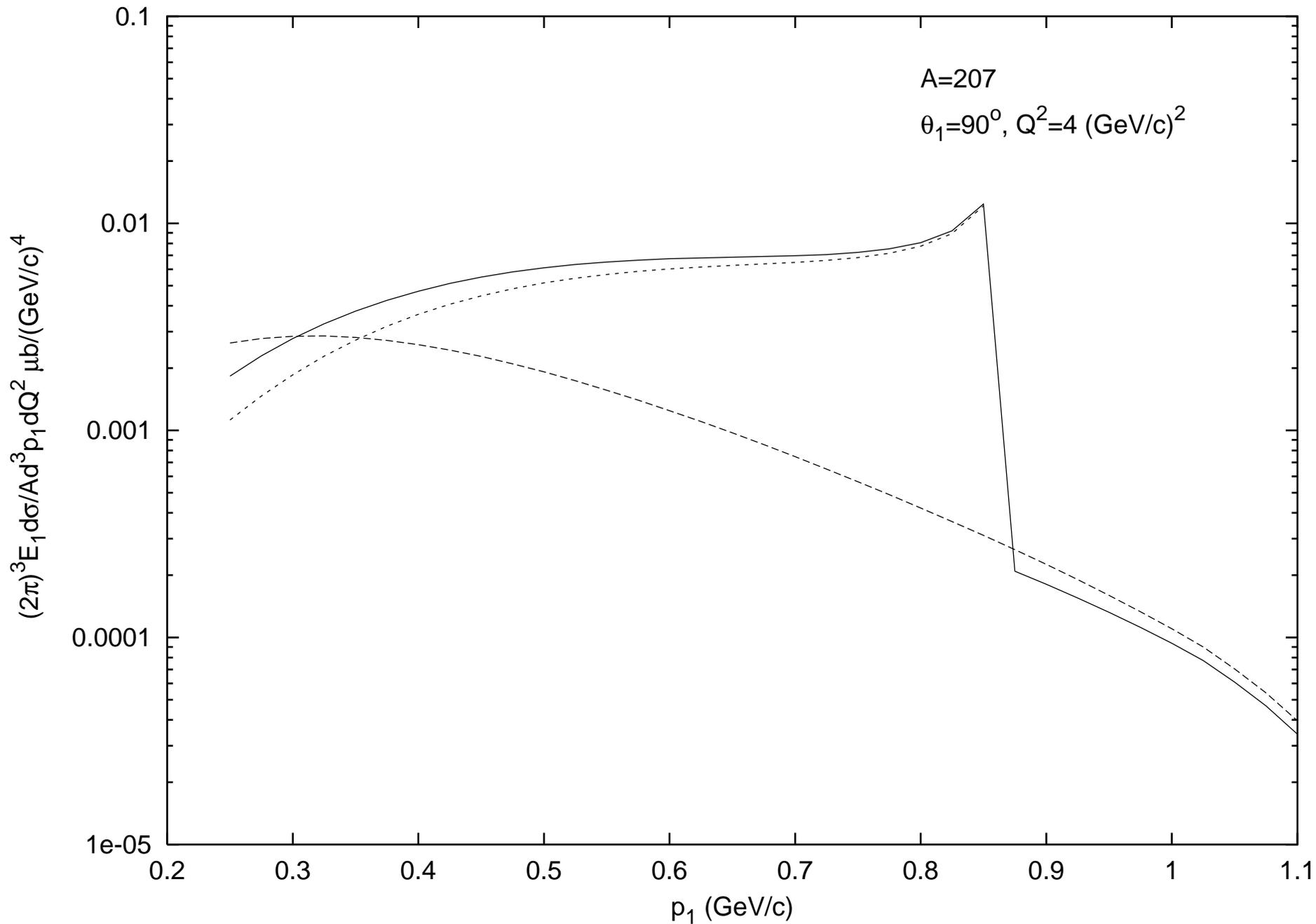
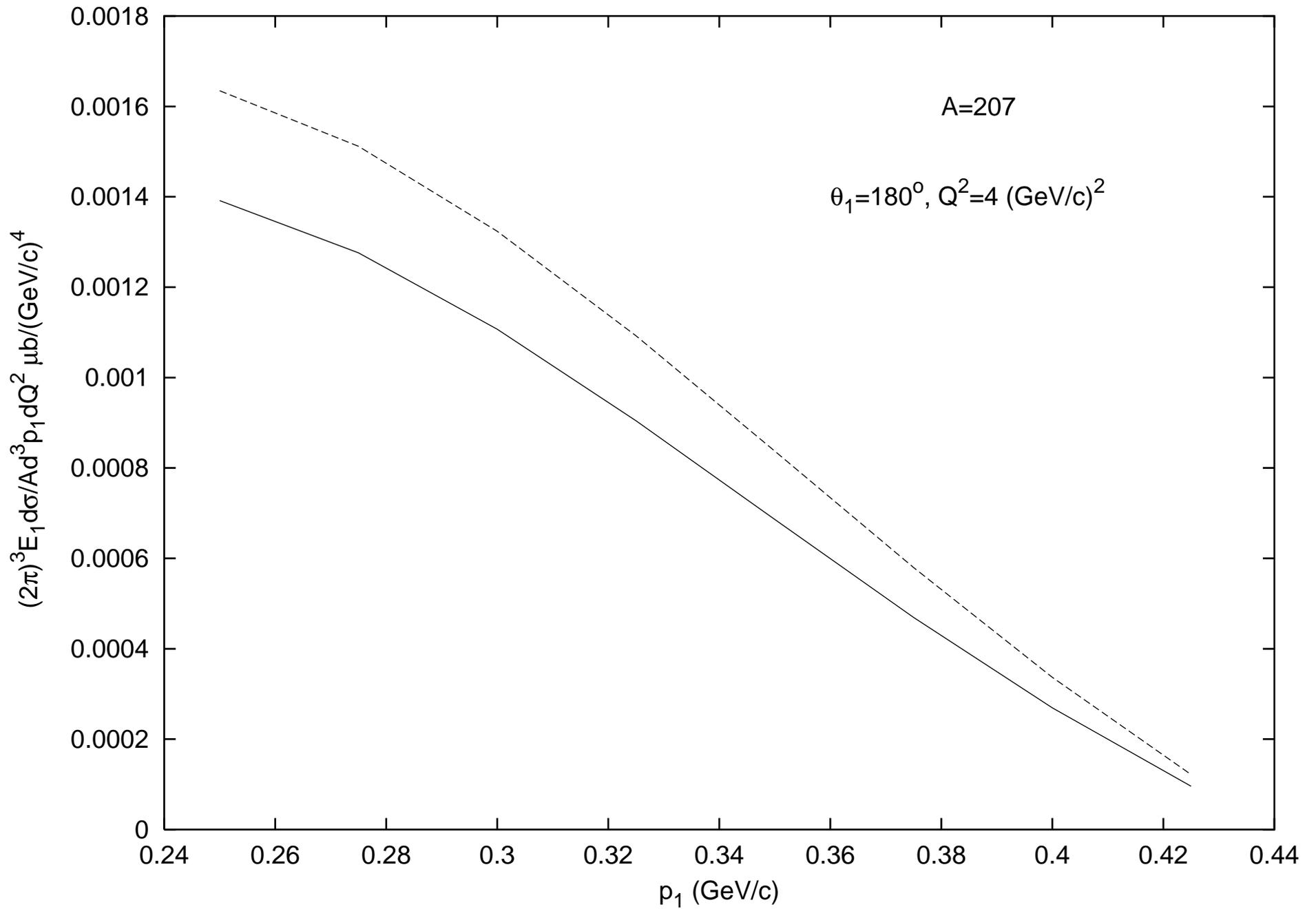
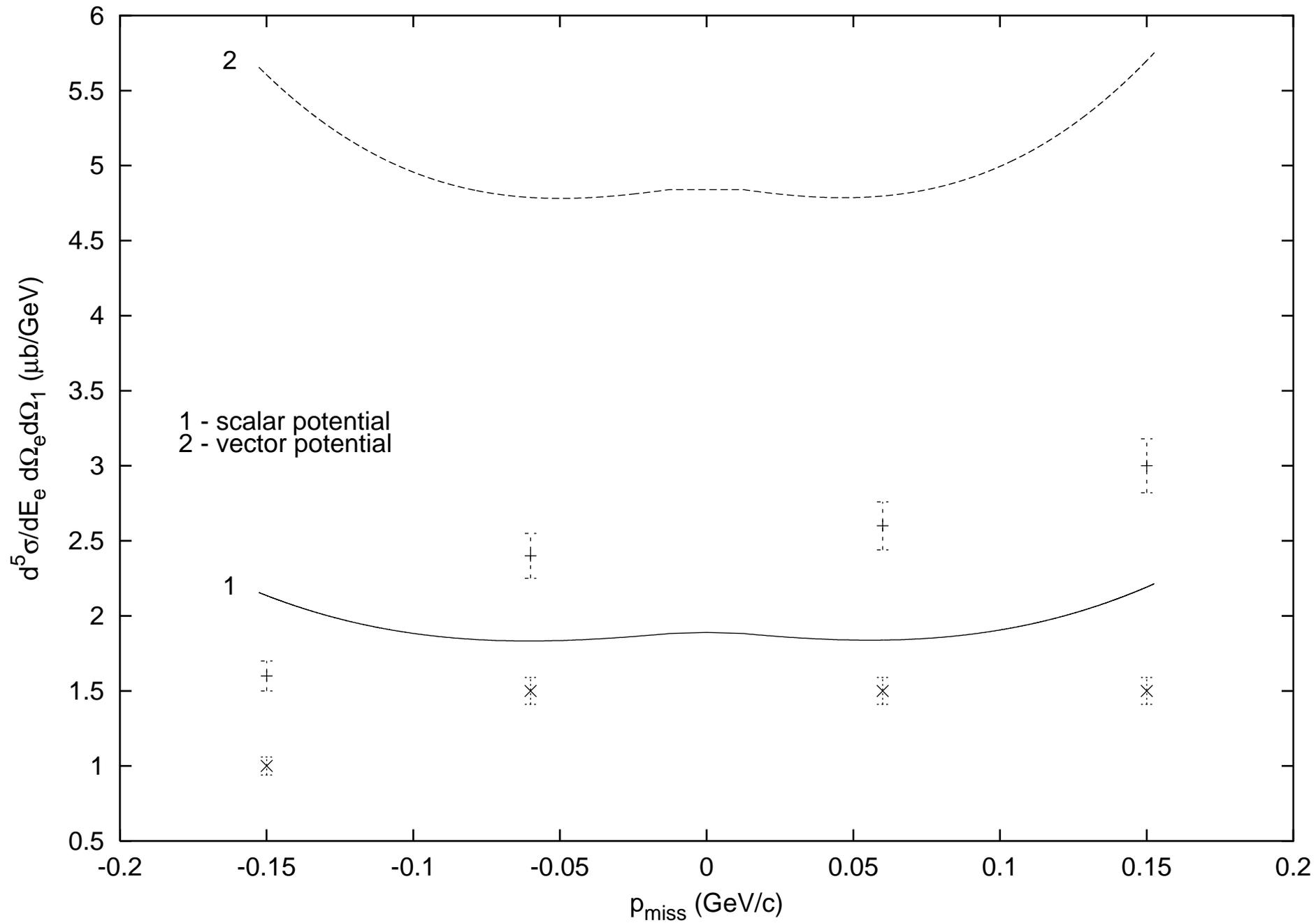
\end{document}